\documentclass[fleqn,usenatbib]{mnras}

\usepackage{newtxtext,newtxmath}
\newcommand\Msun{M$_\odot$~}
\newcommand\Lsun{L$_\odot$~}
\newcommand\Rsun{R$_\odot$~}
\newcommand\swiftxrs{SWIFT\,J0850.8-4219}

\usepackage[T1]{fontenc}

\DeclareRobustCommand{\VAN}[3]{#2}
\let\VANthebibliography\thebibliography
\def\thebibliography{\DeclareRobustCommand{\VAN}[3]{##3}\VANthebibliography}

\usepackage{graphicx}	
\usepackage{amsmath}	

\title[The red supergiant X-ray binary SWIFT\,J0850.8-4219/SRGA\,J085040.6-421156]{Infrared spectroscopy of SWIFT\,J0850.8-4219: Identification of the second red supergiant X-ray binary in the Milky Way}

\author[K. De et al.]{
Kishalay De,$^{1,2}$\thanks{E-mail: kde1@mit.edu}
Fiona A. Daly,$^{1}$
and
Roberto Soria$^{3,4,5}$\\
$^{1}$MIT-Kavli Institute for Astrophysics and Space Research, 77 Massachusetts Ave., Cambridge, MA 02139, USA\\
$^{2}$NASA Einstein Fellow\\
$^{3}$College of Astronomy and Space Sciences, University of the Chinese Academy of Sciences, Beijing 100049, China\\
$^{4}$INAF -- Osservatorio Astrofisico di Torino, Strada Osservatorio 20, I-10025 Pino Torinese, Italy\\
$^{5}$Sydney Institute for Astronomy, School of Physics A28, The University of Sydney, Sydney, NSW 2006, Australia\\
}

\date{Accepted XXX. Received YYY; in original form ZZZ}

\pubyear{2015}

\begin{document}
\label{firstpage}
\pagerange{\pageref{firstpage}--\pageref{lastpage}}
\maketitle

\begin{abstract}
High mass X-ray binaries hosting red supergiant (RSG) donors are a rare but crucial phase in massive stellar evolution, with only one source previously known in the Milky Way. In this letter, we present the identification of the second Galactic RSG X-ray binary \swiftxrs. We identify the source 2MASS\,08504008-4211514 as the likely infrared counterpart with a chance coincidence probability $\approx 5 \times 10^{-6}$. We present a $1.0 - 2.5\,\mu$m spectrum of the counterpart, exhibiting features characteristic of late-type stars and an exceptionally strong He I emission line, corroborating the identification. Based on i) the strength of the $^{12}$CO(2,0) band, ii) strong CN bandheads and absent TiO bandheads at $\approx 1.1$\,$\mu$m and iii) equivalent width of the Mg\,I $1.71\,\mu$m line, we classify the counterpart to be a K3$-$K5 type RSG with an effective temperature of $3820 \pm 100$\,K, located at a distance of $\approx 12$\,kpc. We estimate the source X-ray luminosity to be $(4 \pm 1) \times 10^{35}$\,erg\,s$^{-1}$, with a hard photon index ($\Gamma < 1$), arguing against a white dwarf accretor but consistent with a magnetized neutron star in the propeller phase. Our results highlight the potential of systematic NIR spectroscopy of Galactic hard X-ray sources in completing our census of the local X-ray binary population.
\end{abstract}




\section{Introduction}
Symbiotic X-ray binaries (SyXRBs) are a rare class of X-ray binaries consisting of a neutron star (NS) or black hole (BH) accreting from the wind of a very late-type giant companion \citep{Davidsen1977}. There are only 14 known SyXRBs in the Milky Way \citep{Yungelson2019, De2022b} compared to the hundreds of other known classes of X-ray binaries \citep{Liu2007}. Given their short-lived lifetimes as well as rare formation pathways (e.g. \citealt{Kuranov2015, Hinkle2006}), population synthesis models predict that there are only $\approx 50$ active SyXRBs in the Galaxy. Nearly all known SyXRBs consist of a low mass giant donor; the SyXRB Scutum\,X-1 was proposed to be the first known SyXRB with a red supergiant (RSG) companion \citep{Kaplan2007}, although the source has been recently confirmed to host the first confirmed Mira variable donor in a XRB \citep{De2022b}. \citet{Hinkle2020} recently showed that the late-type donor in the SyXRB 4U\,1954$+$31 was long mis-classified as a low mass giant; instead the infrared luminosity and spectra are consistent with a RSG donor.

4U\,1954+31 remains as the only known RSG X-ray binary in the Milky Way. With a $\approx 7-15$\,\Msun donor, this source is the only high mass X-ray binary (HMXB) with a late-type donor, compared to the $>100$ known HMXBs with blue supergiant donors \citep{Chaty2018}, reflecting the extremely short-lived lifetime of the RSG phase. In addition, the survival of the binary up to the SyXRB phase requires fine-tuned conditions to ensure that the supernova (SN) that created the compact object had sufficiently low ejecta mass and kick velocity to prevent the disruption of the binary. Indeed, \citet{Hinkle2020} invoked a low mass core-collapse SN from an ultra-stripped progenitor \citep{Tauris2015, De2018} to explain the existence of the RSG in 4U\,1954+31. Following the lifetime of the RSG, the final fate of such systems involve the likely creation of a double NS binary, which may merge producing detectable gravitational waves if the orbital period of the resulting double NS binary is short enough \citep{Tauris2006, Tauris2017}.

As such, the identification of every such system in the Galaxy is central to our understanding of these rare evolutionary pathways in massive stellar evolution and our interpretation of the gravitational wave universe. However, the faint X-ray luminosity of wind-accreting SyXRBs, together with their locations in obscured regions of the Galactic plane render identification and detailed characterization of their optical/infrared counterparts challenging. The advent of wide-area, sensitive hard X-ray surveys, together with sensitive near-infrared (NIR) spectrographs on moderate aperture telescopes to perform characterization of the bright infrared counterparts of SyXRBs is beginning to open up a new era in population studies of these unique systems. Here, we present the spectroscopic identification of a new Galactic RSG X-ray binary \swiftxrs, confirmed as part of an ongoing program for systematic infrared spectroscopy of unidentified Galactic plane X-ray sources. 

\section{Observations}

\subsection{Hard X-ray identification and localization}

The hard X-ray source \swiftxrs\ was reported in the 105-month all-sky {\it Swift} Burst Alert Telescope (BAT) catalog \citep{Oh2018}. It was also detected in the soft X-ray band, in two pointed observations with the {\it Swift} X-Ray Telescope (XRT; \citealt{Burrows2005}) done on 2011 December 22 and 24. 
An X-ray source at a very similar location was also detected with signal-to-noise ratio of 5.7 in the first data release \citep{Pavlinsky2021} of the {\it Spektrum-Roentgen-Gamma} ({\it SRG}) Astronomical Roentgen Telescope -- X-ray Concentrator (ART-XC) all-sky survey. The ART-XC detection (catalogued as SRGA\,J085040.6$-$421156) was based on data collected between 2019 December and 2020 December. The nominal positional uncertainty of the ART-XC source is $\lesssim 15$\,\arcsec, which makes it perfectly consistent with the {\it Swift}/XRT position.

\begin{figure}
    \centering
    \includegraphics[width=0.49\textwidth]{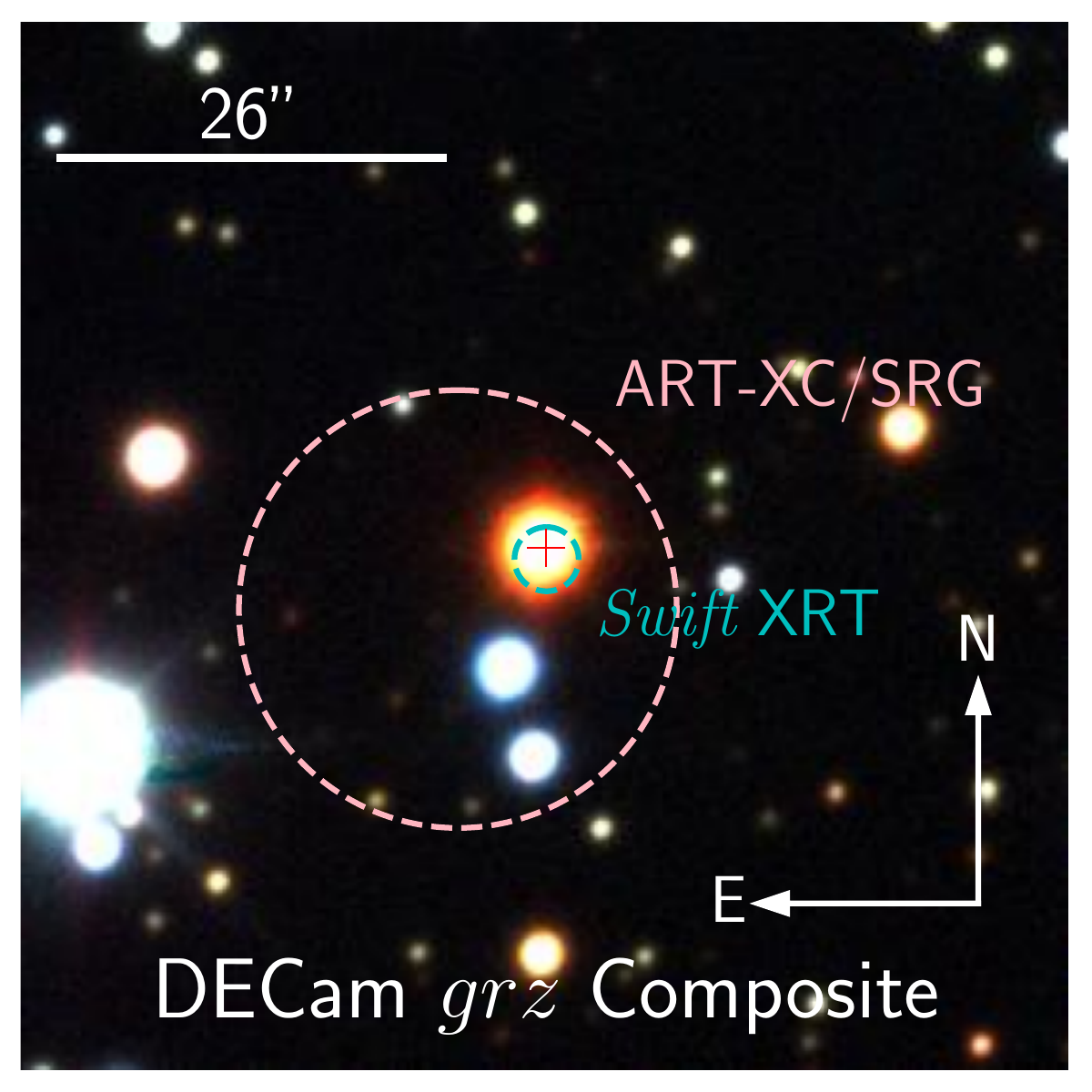}
    \caption{Optical color composite ($grz$ bands) image of the localization region of SWIFT\,J0850.8-4219 from the DECam Galactic Plane Survey. The image shows the 90\% localization region for the X-ray source from the ART-XC catalog and follow-up with Swift XRT. The image is centered on the proposed NIR counterpart 2MASS\,0850, shown with the red cross-hair.}
    \label{fig:localization}
\end{figure}

We re-processed the two public archival {\it Swift}/XRT observations of the field with the online XRT products tool \citep{Evans2009} to derive an enhanced position using stars in the {\it Swift} UltraViolet and Optical Telescope (UVOT; \citealt{Goad2007}) field of view. The best-fit position (J2000 coordinates) is $\alpha = 08^{h}$50$^{m}$40$^{s}$.08, $\delta = -42^\circ 11$\arcmin\,$52\farcs{3}$, with an uncertainty of $2\farcs{2}$. The accurate XRT localization enables the identification of a possible bright NIR counterpart in the 2MASS catalog (2MASS\,08504008$-$4211514; hereafter 2MASS\,0850), located $\approx$0$\farcs{9}$ from the best-fit XRT position. 
In Figure \ref{fig:localization}, we show an optical color composite image of the localization region using archival images from the DECam Galactic plane survey \citep{Schlafly2018}.

The proposed NIR counterpart is reported in the 2MASS catalog \citep{Skrutskie2006} with NIR magnitudes $J = 9.53 \pm 0.02$\,mag, $H = 8.27 \pm 0.03$\,mag and $Ks = 7.77 \pm 0.02$\,mag.

Following \citet{Kaplan2007}, we use the 2MASS catalog to estimate a spatial density of $\approx 2\times 10^{-6}$ arcsec$^{-2}$ for sources brighter than $K_s = 8.0$ in a $10$\,arcmin radius around the source. This suggests a random probability of $\approx 5 \times 10^{-6}$ for the X-ray source to be spatially coincident within $\approx 0\farcs{9}$ of the bright IR source. We therefore conclude that 2MASS\,0850 is the likely IR counterpart of \swiftxrs.

\subsection{SOAR/TSpec Observations}

\begin{figure*}
    \centering
    \includegraphics[width=0.9\textwidth]{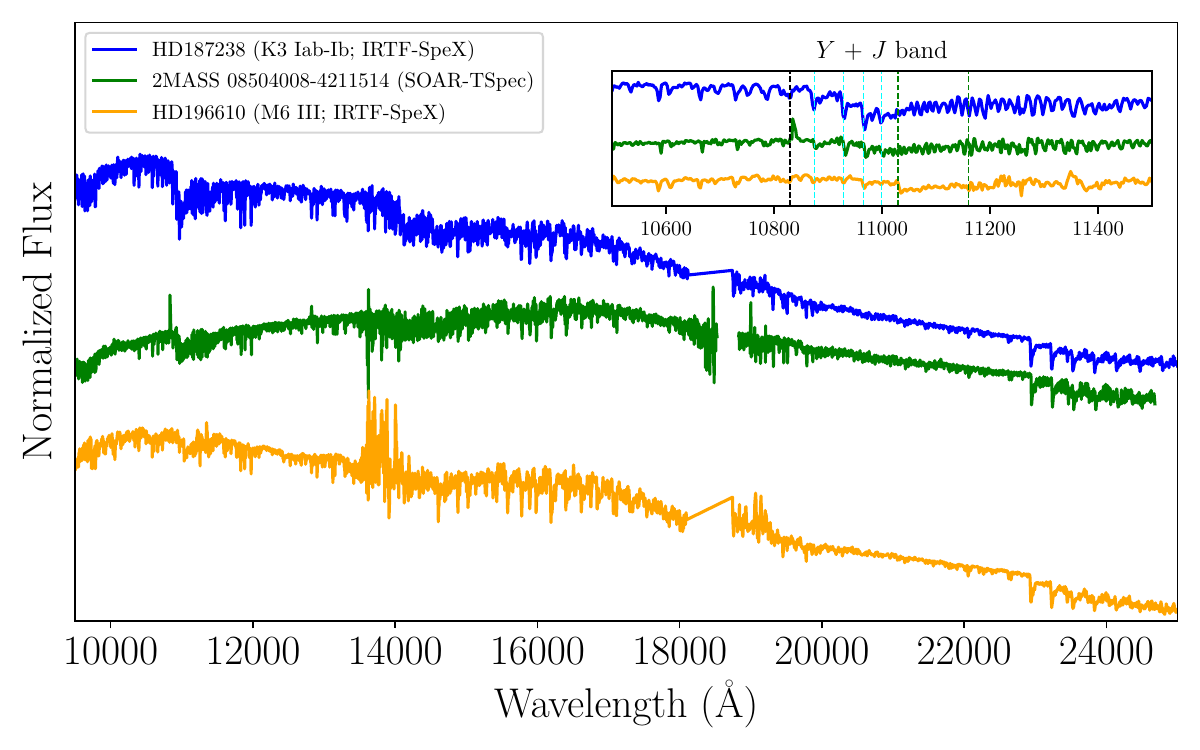}
    \caption{NIR spectrum of 2MASS\,0850, the proposed infrared counterpart of \swiftxrs, compared to late-type stars in the IRTF spectral library \citep{Rayner2009}. We show comparisons against the two possible classifications of the source -- a M6-III giant or a K3 supergiant. The inset shows the spectra zoomed into the region covering $Y$ and $J$ band, highlighting the He\,I\,10830\,\AA feature (black dotted line), strong CN lines (cyan dotted line) seen in supergiants and the TiO bandhead (green dotted line) seen only in late-type giants (but not in 2MASS\,0850). The continuum shape has not been corrected for foreground extinction, and hence the redder continuum of 2MASS\,0850 reflects the larger foreground extinction to the source.}
    \label{fig:spec}
\end{figure*}

On UT 2022-05-12, we obtained a NIR spectrum of 2MASS\,0850 using the TripleSpec spectrograph \citep{Schlawin2014} on the 4.1\,m Southern Astrophysical Research Telescope (SOAR; program 2022A-881609, PI: De). The data were acquired as a series of dithered exposures with ABBA dither pattern, amounting to a total exposure time of 240\,s. We also obtained exposures of the A0V telluric standard HIP\,45079. The data were reduced using the \texttt{spextool} package \citep{Cushing2004} followed by telluric correction and flux calibration using the \texttt{xtellcor} package \citep{Vacca2003}. Figure \ref{fig:spec} shows the $1.0 -2.5\,\mu$m spectrum of 2MASS\,0850. The spectrum exhibits a red continuum peaking in $H$-band together with deep absorption features of the CO first overtone starting at $\approx 2.29\,\mu$m, characteristic of late K-M giants and supergiants. In $Y$-band, we identify a very strong emission line of the He\,I 10830 line, with an equivalent width of $-6.2 \pm 0.1$\,\AA. The strong emission line is indicative of a hot source (i.e. a compact object) ionizing the wind of the star since isolated late-type giants exhibit much weaker He\,I lines ($|EW| \ll 1$\,\AA; \citealt{Dupree2009}), confirming the association of the infrared source with the hard X-ray counterpart. 

\section{Analysis}

\subsection{Spectroscopic diagnostics}

We use equivalent widths (EWs) measured from the NIR spectrum of 2MASS\,0850 to constrain the spectral type and luminosity class of the donor star in \swiftxrs. We use the wavelength regions defined by \citet{Messineo2021} to measure the EWs of the spectral features. The equivalent width of the $^{12}$CO(2,0) bandhead is $35.8 \pm 0.8$\,\AA. We clearly detect absorption features of Na\,I (at $2.207\,\mu$m) and Ca\,I (at $2.263\,\mu$m) in $K$-band, with equivalent widths of $2.2 \pm 0.3$\,\AA\ and $3.1 \pm 0.3$\,\AA\ respectively. Together we measure the $\log [EW({\rm CO})/(EW({\rm Na})+EW({\rm Ca}))]$ to be $\approx 0.83$, squarely pointing to the classification of this source as a late-type giant or supergiant, as late-type dwarfs occupy a different parameter space in this ratio \citep{Ramirez1997}.

If we assume the source to be a giant as seen in most other SyXRBs, the relationship between the $^{12}$CO(2,0) EW and effective temperature in \citealt{Ramirez1997} would suggest a spectral type of $\approx$M5$-$M6, corresponding to an effective temperature of $\approx 3500$\,K. In Figure \ref{fig:spec}, we show a comparison of the full and $Y+J$-band spectrum of a M6-III giant in the IRTF spectral library \citep{Rayner2009} to 2MASS\,0850. As shown, there are striking differences in features that are known to distinguish giants and supergiants \citep{Messineo2021} -- i) the presence of strong CN features in 2MASS\,0850, but weak in the M6-III template, and ii) the absence of a strong TiO bandhead at $\approx 1.1\,\mu$m in 2MASS\,0850, clearly present in the M6-III template and ubiquitously seen in giants of this spectral type.

As supergiants occupy a different relationship between the CO absorption strength and spectral type (owing to their lower surface gravity, which strengthens the CO band for a given temperature), we use the calibration of \citet{Blum2003} to estimate the corresponding effective temperature and spectral type. We measure the CO index, as defined in \citet{Blum2003}, to be $19.27 \pm 0.05$ in 2MASS\,0850. The corresponding effective temperature would be $3820 \pm 10$\,K. \citet{Messineo2021} show that effective temperatures derived from infrared diagnostics typically have a systematic scatter of $\approx 100$\,K from those derived from optical spectroscopic features (the TiO bandhead; \citealt{Dorda2016a, Dorda2016b}), and we therefore adopt the temperature to be $3820 \pm 100$\,K. The corresponding spectral type is $\approx$K3$-$K5 based on the temperature scale of \citet{Levesque2005}.

\begin{figure*}
    \centering
    \includegraphics[width=0.45\textwidth]{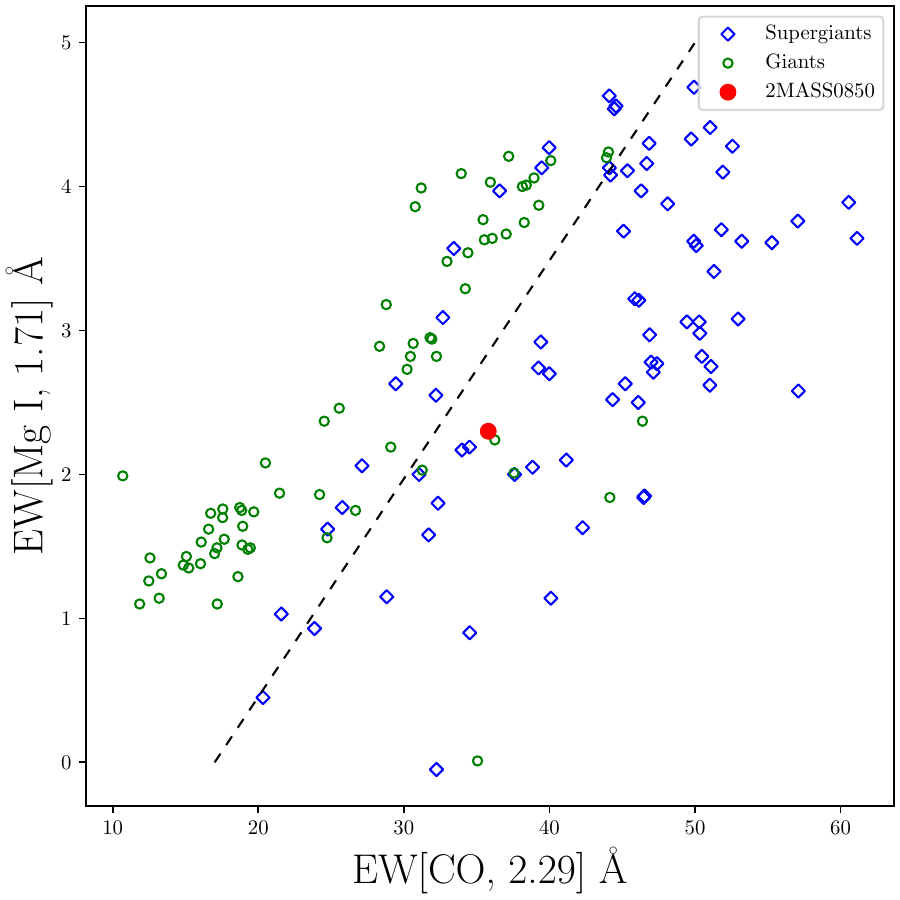}
    \includegraphics[width=0.45\textwidth]{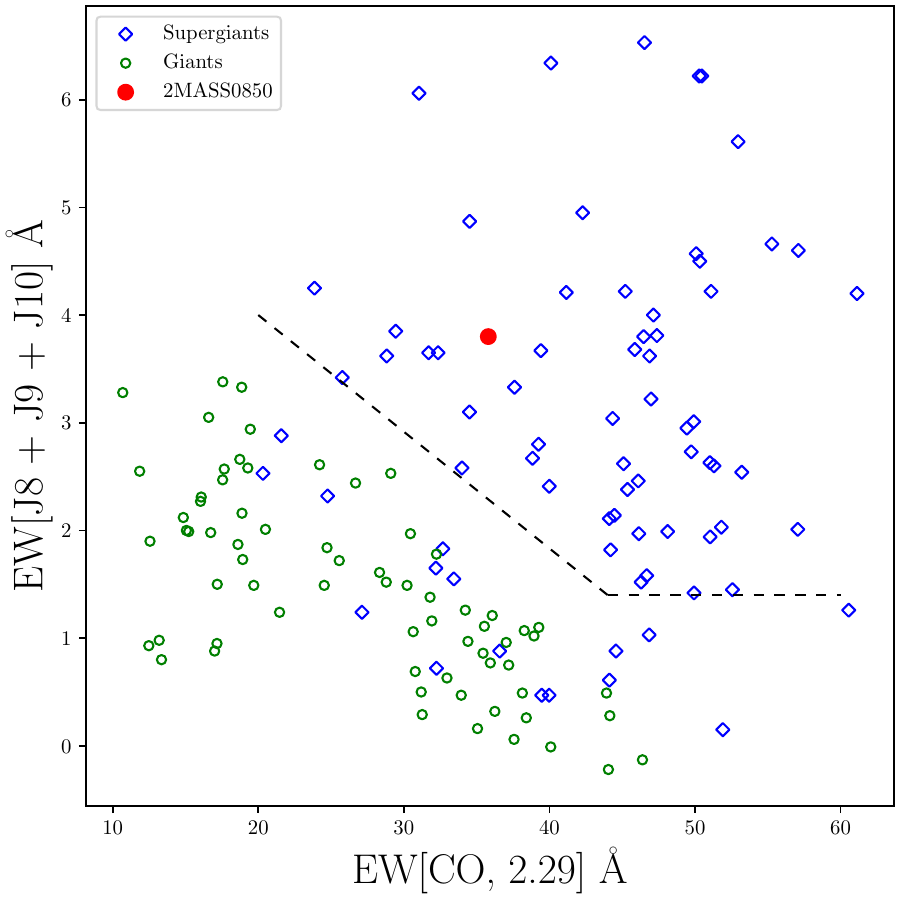}
    \caption{Comparison of the equivalent widths of 2MASS\,0850 to giants and supergiants presented \citet{Messineo2021}. The left plot shows the position of the source in the diagram of Mg I against CO, with the black dashed line separating the regions occupied by giants and supergiants (supergiants occupy the region to the right of the line) as proposed by \citet{Messineo2021}. The right plot shows the same in the combination of the \texttt{J8 + J9 + J10} indices against CO, with supergiants occupying the region above the black dashed line.}
    \label{fig:ewindices}
\end{figure*}

In Figure \ref{fig:spec}, we also show a comparison of the spectrum of a K3 supergiant to 2MASS\,0850 showing better consistency in the spectral features. To quantify these qualitative differences, we turn to the infrared spectral indices presented by \citet{Messineo2021}. They show that giants and supergiants can be differentiated based on their Mg\,I\,$1.71\,\mu$m and CO\,$2.29\,\mu$m equivalent widths. We measure the Mg\,I\,$1.71\,\mu$m EW to be $2.3 \pm 0.2$\,\AA, placing this source in the region occupied by supergiants (Figure \ref{fig:ewindices}). We further measure the \texttt{J8}, \texttt{J9} and \texttt{J10} indices defined in \citet{Messineo2021}, which have been used to detect CN absorption that is ubiquitously seen in RSGs, but weak or absent in giants. We find that the source again lies in the region squarely occupied only by supergiants in the phase space of \texttt{J8+J9+J10} (measured to be $\approx 3.8 \pm 0.2$\,\AA) against CO EW (Figure \ref{fig:ewindices}). Together, we classify 2MASS\,0850 to be a RSG with a spectral type of $\approx$K3$-$K5.

\subsection{Distance, extinction and system paramters}

The integrated Galactic extinction along the line of sight to 2MASS\,0850 is $A_V \approx 7.0$\,mag \citep{Schlafly2011}, and therefore sufficiently large to affect the brightness and color of the source. In order to estimate the distance and absolute luminosity of 2MASS\,0850, we take the typical NIR colors of K3 RSGs of $J - K \approx 0.7$\,mag \citep{Messineo2019} together with the observed 2MASS colors to obtain a foreground extinction of $A_V \approx 5.3$\,mag. The reported parallax of the source in Gaia DR3 \citep{Gaia2023} is $\approx 0.08 \pm 0.01$\,mas, corresponding to a distance range of $\approx 11-14$\,kpc. Correcting for foreground extinction, we estimate the absolute magnitude of the source to be in the range $M_K \approx -8.0$ to $\approx -8.5$\,mag, strikingly consistent with the average absolute magnitude of K3 RSGs ($M_K \approx -8.2$\,mag; \citealt{Messineo2019}). Adopting $12$\,kpc as the distance (corresponding to $M_K = -8.2$\,mag), the corresponding average bolometric luminosity and stellar radius (assuming a blackbody stellar surface) are $\approx 1.7 \times 10^4$\,\Lsun ($\log L_{\rm bol}/L_\odot \approx 4.23$) and $\approx 300$\,\Rsun respectively. 

\subsection{X-ray properties}
\begin{figure}
\hspace{-0.5cm}
    \includegraphics[height=1.05\linewidth, angle=270]{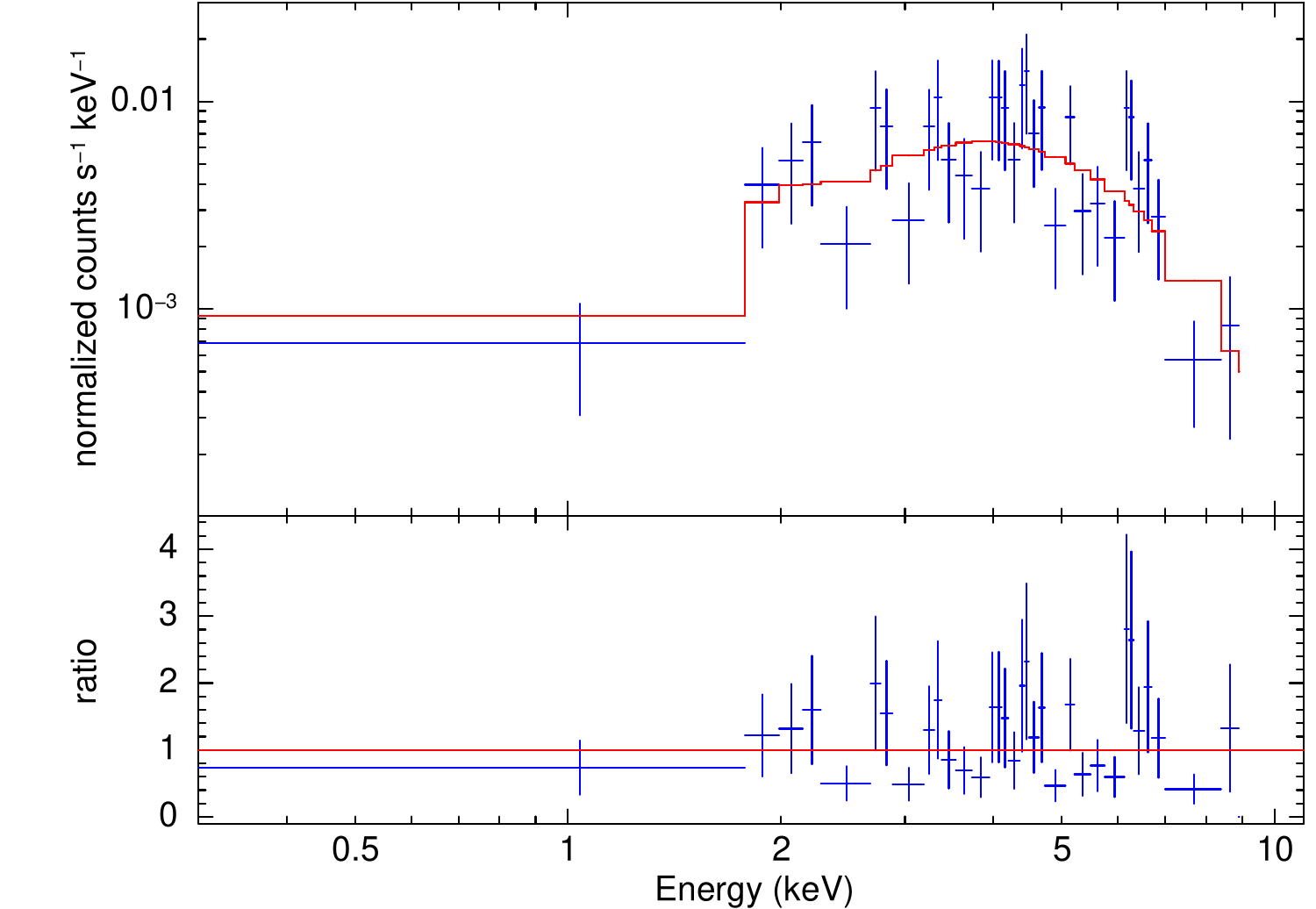}
    \caption{{\it Swift}/XRT spectrum from the combined datasets of 2012 December, with best-fitting power-law model and data/model ratios (see Table  \ref{tab:swift_xrt}). The data have been rebinned to a signal-to-noise ratio $>$1.8 for plotting purposes. }
    \label{fig:swift_xrt}
\end{figure}

We stacked the data from the two {\it Swift}/XRT observations and built a spectrum with the online XRT products tool \citep{Evans2009}. The combined exposure time is 4.8 ks. We fitted the 0.3--10 keV spectrum with {\sc xspec}, version 12.12.0 \citep{arnaud96}. Because of the limited number of counts (only $\approx$130), we could only use a simple power-law model with photoelectric absorption ({\it tbabs} $\times$ {\it pow} in {\sc xspec}). We grouped the spectrum to a minimum of 1 count per bin and used the {\tt cstat} fit statistic \citep{cash79}. We calculated the absorbed and unabsorbed flux and luminosity with the {\it cflux} convolution model. The best-fitting parameters are listed in Table \ref{tab:swift_xrt}, and the spectral fit with its data/model ratio is illustrated in Figure \ref{fig:swift_xrt}. 

The unabsorbed 0.3--10 keV luminosity from the XRT is $L_{0.3-10} \approx (1.2 \pm 0.2) \times 10^{35}$ erg s$^{-1}$. This is consistent with the results reported in the ART-XC catalog \citep{Pavlinsky2021} over the 4--12 keV band: $f_{4-12} \approx 8.8^{+3.2}_{-2.6} \times 10^{-12}$\,erg\,cm$^{-2}$\,s$^{-1}$; $L_{4-12} \approx 1.5^{+0.5}_{-0.2} \times 10^{35}$\,erg\,s$^{-1}$. It suggests that the source has not varied significantly during the nine years between the {\it Swift}/XRT and {\it SRG}/ART-XC observations. In addition, a hard X-ray flux $f_{14-195} \approx 12.1^{+2.0}_{-3.0} \times 10^{-12}$\,erg\,cm$^{-2}$\,s$^{-1}$, corresponding to a luminosity $L_{14-195} \approx 2.1^{+0.3}_{-0.5} \times 10^{35}$\,erg\,s$^{-1}$, was reported in the BAT catalog  \citep{Oh2018} over the 14--195 keV band. In summary, we estimate a total X-ray luminosity $L_{\rm X} \approx (4 \pm 1)  \times 10^{35}$\,erg\,s$^{-1}$.

The high X-ray luminosity exceeds that seen in accreting white dwarf systems ($\lesssim 10^{34}$\,erg\,s$^{-1}$; \citealt{Mukai2017}), which rules out a Cataclysmic Variable identification. At the same time, the unusually hard photon index $\Gamma < 1$ in the 2--10 keV band (Table \ref{tab:swift_xrt}) is inconsistent with the typical range of photon indices observed in the low/hard state of stellar-mass BHs, that is $1 < \Gamma \lesssim 2$ at similar luminosities \citep{remillard06,plotkin13,yang15,liu19}. The photon index is also much lower than those observed in low-mass X-ray binaries powered by weakly magnetized NSs \citep{wijnands15}. It is instead perfectly consistent \citep{reig13,farinelli16} with the photon indices observed from accreting, strongly magnetized NSs (X-ray pulsars) in HMXBs. The spatial association with a bright 2MASS star further supports the scenario of a young, strongly magnetized NS accreting from a RSG. Thus, we conclude that \swiftxrs\ is most likely a newly identified Galactic SyXRB with a RSG donor.

\subsection{Supergiant wind and accretion rate }
A further test on the SyXRB scenario is whether the wind from the relatively faint RSG 2MASS\,0850 can provide enough matter to the accreting NS to explain the system's X-ray luminosity. The mass accretion rate onto the NS can be approximated as 
\begin{equation}
    \dot{M} \approx 2.5 \pi G^2 M_{\rm ns}^2 \rho_{\rm w} v_{\rm rel}^{-3} 
\end{equation}
\citep{bondi44}, where $M_{\rm ns} \approx 1.4M_{\odot}$ is the mass of the NS, $\rho_{\rm w}$ is the density of the RSG wind at the orbit of the NS, and $v_{\rm rel} \approx \left(v_{\rm w}^2 + v_{\rm K}^2   \right)^{1/2}$ is the relative velocity of stellar wind and compact object, function both of the intrinsic wind speed $v_{\rm w}$ and of the Keplerian orbital speed $v_{\rm K}$ of the NS. The wind density is related to the RSG mass loss rate $\dot{M}_{\rm w}$ as 
\begin{equation}
    \dot{M}_{\rm w} = 4 \pi a^2 \rho_{\rm w}(a) v_{\rm w}(a)
\end{equation}
where $a$ is the binary separation. The wind speed depends on metallicity and other properties of the RSG star, but is generally observed to be $\sim$10--30 km s$^{-1}$ \citep{vanloon01,goldman17}. That also implies that the dominant contribution to the relative speed comes from the orbital motion of the NS, rather than from the wind speed (the opposite as in blue supergiant systems). For simple order-of-magnitude estimates, we can assume that the mass of the NS is much lower than the mass of the RSG donor ($M \sim 10$--20$M_{\odot}$), and we can take the simple case of a circular orbit. Then, the Keplerian speed $v_{\rm K} \approx100 \left(M/15M_{\odot}\right)^{1/2}\, \left(a/300R_{\odot}\right)^{-1/2}$ km s$^{-1}$. For plausible binary separations $a \lesssim 10^3 R_{\odot}$, corresponding to orbital periods $P \lesssim 10^3$ d, we can also approximate  $v_{\rm rel} \approx v_{\rm K}$.

This provides a simple estimate for the fraction of RSG wind accreted by the NS, from the ratio of equations (1) and (2):
\begin{equation}
    \frac{\dot{M}}{\dot{M}_{\rm w}} \approx 0.03 \left(\frac{M_{\rm ns}}{1.4M_{\odot}} \right)^2
    \left(\frac{M}{15M_{\odot}} \right)^{-3/2}
    \left(\frac{a}{300R_{\odot}} \right)^{-1/2}
    \left(\frac{v_{\rm w}}{20{\rm {~km~s}^{-1}}} \right)^{-1}
\end{equation}

Determining the RSG mass loss rate is notoriously one of the critically unsolved problems of stellar evolution. For a given observed luminosity, model estimates and empirical constraints of $\dot{M}_{\rm w}$ can vary by three orders of magnitude, from $\sim$a few $10^{-9}$ $M_{\odot}$ yr$^{-1}$ to a few $10^{-6}$ $M_{\odot}$ yr$^{-1}$ 
\citep{vanloon05,ekstrom12,meynet15,beasor20,humphries20,wang21,massey23}. For our purpose, we take a value $\dot{M}_{\rm w} \sim 10^{-7} M_{\odot}$ yr$^{-1}$, in the middle of the range of published values, for a luminosity $L_{\rm bol} \approx 10^{4.23} L_{\odot}$. This corresponds to an accretion rate $\dot{M} \sim 2 \times 10^{17}$ g s$^{-1}$ for the fiducial parameters in Equation (3), and a corresponding luminosity $L_{\rm X} \sim 3 \times 10^{37}$ erg s$^{-1}$ (for efficient accretion onto the NS surface, $\eta \approx 0.2$). Thus, our estimate shows that the wind of a RSG donor can indeed provide sufficient mass accretion to explain the X-ray luminosity of \swiftxrs. Even with the lower mass loss rates ($\dot{M}_{\rm w} \sim 10^{-8} M_{\odot}$ yr$^{-1}$) proposed by \cite{beasor20}, the accretion rate is more than enough to produce the observed luminosity.

The next challenge of our scenario is to explain why the observed $L_{\rm X}$ is {\it only} $\sim$10$^{35}$ erg s$^{-1}$ rather than $\sim$10$^{37}$ erg s$^{-1}$. A similar issue was discussed by \citet{Hinkle2020} in relation to the relatively low luminosity of the SyXRB 4U\,1954$+$31. In that case, the explanation suggested by \citet{Hinkle2020} was that the system had a very large binary separation, with a period of (at least) several years, so that $\dot{M}/\dot{M}_{\rm w}$ is much reduced. 
For our adopted mass loss rate $\dot{M}_{\rm w} \sim 10^{-7} M_{\odot}$ yr$^{-1}$, and under the assumption of efficient accretion onto the NS surface, we would need a ratio $\dot{M}/\dot{M}_{\rm w} \sim 4 \times 10^{-4}$ to explain the low observed luminosity.

From Equations (1--2), we can recalculate the ratio $\dot{M}/\dot{M}_{\rm w}$ for the more general case of $v_{\rm w} \gtrsim v_{\rm K}$, a regime valid at large binary separations. We estimate that the binary separation would have to be $a \sim 25,000 R_{\odot}$ to justify the low observed luminosity. This would imply an unlikely binary period $P \approx 10^{11}$ s $\approx 10^5$ d $\approx 300$ yr. Extreme fine tuning would be required to produce a system with such marginal binding energy. Even for a lower mass loss rate $\dot{M}_{\rm w} \sim 10^{-8} M_{\odot}$ yr$^{-1}$, we would still need $a \sim 8,000 R_{\odot}$ and $P \approx 50$ yr. Thus, we regard this scenario as implausible.

Instead, we suggest that a more likely explanation is that \swiftxrs\ is in the propeller state \citep{illarionov75,ghosh79,stella86}, in which most of the accretion flow is stopped at the magnetospheric radius. The minimum accretion luminosity below which a magnetized NS is expected to switch to the propeller state is 
\begin{equation}
L_{\rm acc,m} \approx 1.3 \times 10^{37}  \, B_{12}^2 \, P_{\rm s}^{-7/3} \, M_{1.4}^{-2/3} \, R_6^5  \ \ {\mathrm{erg~s}}^{-1}
\end{equation}
\citep{campana18}, where $B_{12}$ is the surface magnetic field in units of $10^{12}$ G, $P_{\rm s}$ is the spin period in seconds, $M_{1.4}$ is the mass of the NS in units of 1.4$M_{\odot}$, and $R_6$ is the NS radius in units of $10^6$ cm. We speculate that the accretor in \swiftxrs\ is a young NS with $B \sim$ a few $10^{12}$ G and $P_{\rm s} \lesssim 1$, so that the accretion rate supplied by the RSG wind is (currently) not enough to keep it in the accretor state.

\begin{table}
\caption{Best-fitting parameters of the {\it Swift}/XRT spectrum of X-1. The model is {\it tbabs} $\times$ {\it powerlaw}, and was fitted with the Cash statistics. Uncertainties correspond to $\Delta C = \pm 2.706$, which is asymptotically equivalent to the 90\% confidence limit in the $\chi^2$ statistics.} 
\vspace{-0.3cm}
\begin{center}  
\begin{tabular}{lc} 
 \hline 
\hline \\[-8pt]
  Model Parameters      &      Values   \\
\hline\\[-9pt]
   $N_{\rm {H}}$   ($10^{22}$ cm$^{-2}$)   &  $ 3.3^{+2.7}_{-1.9}$  \\[4pt]
   $\Gamma$      &  $ 0.3^{+0.7}_{-0.6}$ \\[4pt] 
   $N_{\rm {po}}^a $ 
              &  $ 1.5^{+3.5}_{-1.0}$\\[4pt]
   C-stat/dof     &      $110.0/128$ (0.86) \\[4pt]
   $f_{0.3-10}$ ($10 ^{-12}$ erg cm$^{-2}$ s$^{-1}$)$^b$ 
       & $5.9^{+1.7}_{-1.2}$\\[4pt]
   $L_{0.3-10}$ ($10 ^{35}$ erg  s$^{-1}$)$^c$ 
      & $ 1.2^{+0.2}_{-0.2}$ \\[2pt]
\hline 
\vspace{-0.5cm}
\end{tabular}
\end{center}
\begin{flushleft} 
$^a$: units of $10^{-4}$ photons keV$^{-1}$ cm$^{-2}$ s$^{-1}$ at 1 keV.\\
$^b$: observed flux in the 0.3--10 keV band\\
$^c$: isotropic unabsorbed luminosity in the 0.3--10 keV band, at a distance of 12 kpc.
\end{flushleft}
\label{tab:swift_xrt}
\end{table}

\section{Summary}

In this paper, we present the identification, confirmation and classification of the NIR counterpart to the hard X-ray source \swiftxrs. Using re-analysis of archival X-ray data and new infrared spectroscopic observations, we show
\begin{itemize}
    \item Using archival Swift XRT observations, we localize the X-ray source (with a positional uncertainty of 2.2\arcsec) to be coincident within 0.9\arcsec of the bright NIR source 2MASS\,08504008-4211514. We estimate the chance coincidence probability of the X-ray source being coincident with a NIR source at this brightness to be $\approx 5 \times 10^{-6}$ given the IR source density at this position, supporting the identification of the NIR source as the likely counterpart.
    \item We present a $1.0 - 2.5\,\mu$m NIR spectrum of the counterpart obtained with the SOAR telescope. The spectrum shows typical features of a late-type star together with a strong He\,I emission line (EW $\approx -6.2 \pm 0.1$\,\AA). The strong emission line is not seen in isolated late-type giants, and argues for the presence of a hot ionizing source near the giant, thereby confirming the identification of the counterpart.
    \item Using the CO bandhead at $2.29\,\mu$m to identify the spectral type of the star, we qualitatively show that the spectrum is inconsistent with a giant of the spectral type ($\approx$M5$-$M6) suggested by the EW of the CO bandhead, but is instead consistent with a RSG of spectral type $\approx$K3$-$K5 and effective temperature $3820 \pm 100$\,K. We quantitatively show that the EWs of the CO $2.29\,\mu$m bandhead, the Mg I\,$1.71\,\mu$m absorption feature, and in particular, the strong CN bandheads at $\approx 1.1\,\mu$m are inconsistent with those seen in late-type giants, but are instead consistent with a RSG \citep{Messineo2021}.
    \item Using the inferred spectral type of the RSG from the spectroscopic diagnostics, we estimate the source to be located at a distance of $\approx 12$\,kpc and behind a foreground dust extinction column of $A_V \approx 5.3$\,mag, consistent with the Gaia DR3 parallax of the source.
    \item Analyzing archival X-ray observations from Swift/XRT, SRG/ART-XC and Swift/BAT, we find the X-ray luminosity of the source to be $L_X \approx (4 \pm 1) \times 10^{35}$\,erg\,s$^{-1}$, which together with the hard photon index ($\Gamma < 1$), strongly argues against an accreting white dwarf (which have $L_X \lesssim 10^{34}$\,erg\,s$^{-1}$; \citealt{Mukai2017}) as the compact object. The measurements are instead consistent with the paramters typically seen in accreting, strongly magnetized NSs. 
    \item We show that the wind of the RSG is expected to provide enough material to an accreting NS for reasonable ranges of orbital periods and wind velocities to power the observed X-ray luminosity. We speculate that the $\approx 100\times$ lower observed X-ray luminosity compared to the case of efficient accretion onto the NS can be explained if the accretor is in the propeller state where most of the accretion is stopped at the magnetospheric radius. 
\end{itemize}

Our new identification of \swiftxrs\ as only the second RSG X-ray binary in the Galaxy demonstrates the potential of hard X-ray instruments with high angular resolution localization capabilities to complete our census of the local stellar graveyard and our understanding of massive stellar evolution. Systematic NIR spectroscopy of counterparts identified in future and more sensitive source catalogs from ongoing surveys will provide an unprecedented view into the Galactic X-ray binary populations. 

\section*{Acknowledgements}
We thank E. Kara and R. Simcoe for helpful discussions on the work. We thank M. Messineo for providing the spectral indices used in their paper.  K. D. was supported by NASA through the NASA Hubble Fellowship grant \#HST-HF2-51477.001 awarded by the Space Telescope Science Institute, which is operated by the Association of Universities for Research in Astronomy, Inc., for NASA, under contract NAS5-26555. 
RS acknowledges grant number 12073029 from the National Natural Science Foundation
of China (NSFC).
Based on observations obtained at the Southern Astrophysical Research (SOAR) telescope, which is a joint project of the Minist\'{e}rio da Ci\^{e}ncia, Tecnologia e Inova\c{c}\~{o}es (MCTI/LNA) do Brasil, the US National Science Foundation’s NOIRLab, the University of North Carolina at Chapel Hill (UNC), and Michigan State University (MSU).

\bibliographystyle{mnras}
\bibliography{example} 

\bsp	
\label{lastpage}
\end{document}